\documentstyle[twocolumn,pra,aps,epsfig]{revtex}
\begin{document}
\flushbottom
\draft
\title{Cavity atom optics and the `free atom laser'}
\author{J. Heurich, M. G. Moore, and P. Meystre}
\address{Optical Sciences Center and Department of Physics\\
University of Arizona, Tucson, Arizona 85721\\
\date{June 25, 1999}
\medskip}
\author{\small\parbox{14.2cm}{\small \hspace*{3mm}}}
\maketitle
\begin{abstract}
The trap environment in which Bose-Einstein condensates are generated
and/or stored  
strongly influences the way they interact with light. The situation is analogous to
cavity QED in quantum optics, except that in the present case, one tailors the
matter-wave mode density rather than the density of modes of the optical
field. Just as in QED, for short times, the atoms do not sense the trap
and propagate as in free space.
After times long enough that recoiling atoms can probe the trap environment,
however, the way condensates and light fields are mutually influenced differs
significantly from the free-space situation. We use as an example the 
condensate collective atomic recoil laser, which is the atomic matter-wave 
analog of the  free-electron laser.
\end{abstract}

\section{Introduction}
The manipulation of atomic trajectories by optical fields forms the basis
of atom optics \cite{AdaSigMly94}, a topic of considerable current interest.
Many atom optics experiments proceed by simply reversing the roles
of light and matter from the situation in conventional optics. That is,
the ``optical elements'' used to refract, diffract, reflect, focus, trap,
etc. matter waves are oftentimes --- but not always --- made of light,
which transfers momentum to the atoms to achieve the desired goal. In
many situations, it is sufficient to treat the optical fields as constant
and imposed from the outside. In addition, for low enough densities,
collisions can be ignored.

In general, it is necessary to describe both the Maxwell
field describing light and the Schr\"odinger field describing the atoms
as dynamically coupled. Under certain circumstances, one can then formally
eliminate the dynamics of the atomic field, resulting in effective
interactions
between the light fields. Under a different set of conditions, one can
alternatively eliminate all or part of the electromagnetic field, resulting
in effective atom-atom interactions (collisions). These are the regimes of
{\em nonlinear optics} \cite{She84} and {\em nonlinear atom optics}
\cite{LenMeyWri93,ZhaWalSan94}, respectively. But they
represent limiting cases, where either the atomic or the optical field is
not
dynamically independent, following instead the other field in some
adiabatic manner that allows for its effective elimination. Outside of these
two regimes the atomic and optical fields are dynamically independent. Neither
field is readily eliminated.

In the past, there have been numerous examples where the dynamical coupling
between the optical and matter-wave fields was important. Much of quantum
optics and of laser physics deals with such situations. However, in the
spirit
of atom optics, we wish to concentrate only on those cases where the
center-of-mass motion of the material system is central to the problem.
An early example of such a situation is the free-electron laser (FEL)
\cite{Eli76,HopMeyScu76,Bra90}, where a periodic magnetic field
(the wiggler) and a running wave optical probe field conspire to
spatially modulate the density of a relativistic electron beam. This density
modulation results in an oscillating current which amplifies the
optical field. As a result, the electron density modulation is itself
increased, leading to a runaway amplification process. A similar mechanism
governs the collective atomic recoil laser (CARL), \cite{BonDesNar94} except
that in that case the relativistic electrons are replaced by atoms, and the
periodic magnetic field by an optical pump field. Here, runaway
amplification
results from the stimulated scattering of the pump field off the density
grating imposed on the atomic sample by the combined action of the pump and
probe fields. Similarly to the free-electron laser nomenclature, one would
therefore be justified in calling the CARL a ``free-atom laser.''

Both the FEL and the CARL operate under conditions such that the electrons
or the atoms can be described as classical particles. Indeed, in analogy
with the situation in optics, one can expect that the wave nature of the
particles involved is irrelevant if the associated de Broglie wavelength
is small compared to the characteristic length of the ``optical elements''
involved in their manipulation. For the case of the CARL, for example, this
implies that a classical description of the atomic trajectories is
appropriate only for temperatures large compared to the recoil
temperature.

The situation becomes in many ways much more interesting when
the wave properties of the atoms become essential. Early examples of such
situations were discussed by Meystre et al, \cite{MeySchSte89} who studied
the diffraction of an atom by a quantized field mode, by Haroche et al,
\cite{HarBruRai91} who discussed the possible trapping of an atom in the
vacuum field of a high-Q microwave cavity, and by Englert et al,
\cite{EngSchBar91} who considered the reflection of an ultracold atom at
the entrance of such a cavity. Scully and coworkers subsequently extended
this work to the theory of the micromazer, an ultracold-atom version of the
micromaser.\cite{ScuMeyWal96}

In parallel to these efforts, extensions of the CARL theory into the regime
of ultracold atoms have also been carried out.\cite{MooMey98} They bring to 
light both the impact of atomic diffraction on the threshold 
behavior of the system and the role of the vacuum fluctuations of 
the Schr\"odinger field on the build-up
from noise of the signal. Recently, this work was further extended to
analyze \cite{MooMey99} the superradiant Rayleigh scattering from a 
condensate observed by Ketterle's group.\cite{Ino99}

These brief remarks touch upon just one of the many contributions of Marlan
Scully, yet they clearly illustrate his impact on modern quantum optics. It
is particularly noteworthy that his ``ancient'' work on the FEL and cavity
QED has now gained a second youth, and is central to much of the exciting
work taking place in Bose-Einstein condensation, atom lasers, and the atom
optics of ultracold atom samples.

The present paper discusses some new results in matter-wave optics that
build on the kind of ideas which have been central to Marlan Scully's
research, namely laser physics \cite{SarScuLam74}, amplifier theory, and 
cavity QED \cite{ScuZub97}. The new
twist here is that instead of tailoring the density of modes of the
electromagnetic field, we consider a situation where it is the matter-wave
density of states that is manipulated.

Atomic Bose-Einstein condensates are always formed in traps, a feature that
has of course important implications for many of their properties,
such as e.g. the dependence of the critical temperature on particle
number, the quasi-excitation spectrum, etc. In optical experiments involving
the buildup of condensate momentum side modes, however, one usually considers
a plane-wave expansion rather than the actual matter-wave modes in the
trap. This assumes implicitly that the side modes behave as if in free
space.
The question, then, is to which extent this is appropriate and when a proper
description of the matter-wave modes of the trap is necessary.

The problem at hand is similar to the situation in cavity QED: for
instance, it is known that spontaneous emission always first occurs at the
free space rate, until the wave packet emitted by the atom has a chance to
probe the cavity boundaries and be reflected back to the atom.
\cite{ParStr87,Gie96} This leads to interference effects that can
result either in enhanced or inhibited spontaneous emission, or, for
sufficiently high-$Q$ resonators, to reversible spontaneous emission.
Likewise,
it is expected that matter waves will behave as in free space for times
short compared to their propagation time to the trap boundaries. For
longer times, however, cavity effects are expected to play an important
role. In situations where matter-wave side modes result from an interaction
with
light, the characteristic atomic velocity is the recoil velocity $v_{rec} = 
\hbar k/m$, which is typically of the order of centimeters per second. For
trap
sizes of 100 microns or so, this corresponds to characteristic times of
the order of tens of milliseconds, after which the atoms will have probed
their environment and determined that they are in a trap.

This paper illustrates the role of cavity effects in matter-wave optics on
the example of the ultracold CARL. We first briefly present our model,
and analyze the system's dynamics in the small signal regime
in a way that exposes the differences between the free-space and trap
description of the system. We then compare the instabilities associated
with the two descriptions, showing the appearance of a new kind of
instabilities in the ``cavity matter-wave optics'' analysis.

\section{model}

Our model consists of a Schr\"odinger field of non-interacting
bosonic two-level atoms coupled via the electric-dipole interaction to two
single-mode running wave light fields. We consider only the case where
the detuning between the optical fields and the atomic resonance is many
orders of magnitude larger than the natural linewidth of the
atomic transition. While single photon processes are therefore non-resonant,
the atoms may still undergo two-photon virtual transitions in which their
internal state remains unchanged, but due to recoil may result in a change
in their center-of-mass motion.
In the far-off resonant regime, the excited state population, and therefore
spontaneous emission, may be neglected, and the ground state atomic field
then
evolves coherently under the effective Hamiltonian
\begin{eqnarray}
\hat{\cal H}&=&
\int d^3{\bf r}\hat{\mit\Psi}^\dag({\bf r})
\left[-\frac{\hbar^2}{2m}\nabla^2+V({\bf r})
+\hbar\frac{g^\ast_1g_2}{\Delta}\hat{a}^\dag_1a_2e^{-i{\bf K}\cdot{\bf
r}}\right.
\nonumber\\
&+&\left. \hbar\frac{g^\ast_2g_1}{\Delta}a^\ast_2\hat{a}_1e^{i{\bf
K}\cdot{\bf r}}
\right]\hat{\mit\Psi}({\bf
r})+\hbar(\omega_1-\omega_2)\hat{a}^\dag_1\hat{a}_1,
\label{H}
\end{eqnarray}
where $m$ is the atomic mass, $V({\bf r})$ is the trap potential,
$g_1$ and $g_2$ are the probe and pump coupling coefficients, respectively,
and ${\bf K}={\bf k}_1-{\bf k}_2$ is the difference between the probe and
pump wavevectors. The operator $\hat{a}_1$ is the photon annihilation 
operator for the probe mode, taken in the frame rotating at the pump 
frequency $\omega_2$, hence the energy of the probe mode is the difference
between the probe frequency $\omega_1$ and that of the pump.
The pump is treated classically, and assumed to remain undepleted. Thus
$a_2$ is simply a constant, related to the pump intensity $I_2$
by $|g_2|^2|a_2|^2=d^2I_2/2\hbar^2\epsilon_0c$, where $d$ is the magnitude
of the atomic dipole moment. We remark that we have neglected terms
corresponding to the spatially independent light shift potential. These
terms do not influence the dynamics, as all atoms inside the light field 
undergo the same phase shift. Note that this shift modifies the dispersion
relation for the light fields, hence by ignoring it we are taking the index
of
refraction of the atomic sample to be the same as that of the vacuum.

We assume that the atomic field is initially a Bose-Einstein condensate with
mean number of condensed atoms $N$, and that this condensate is well
described
by a number state so that the initial state of the atomic field may be taken
as
\begin{equation}
|\psi\rangle(t=0)=\frac{1}{\sqrt{N!}}\left(\hat{c}^\dag_0\right)^N|0\rangle,
\label{initial}
\end{equation}
where $|0\rangle$ is the vacuum state, and
\begin{equation}
\hat{c}^\dag_0=\int d^3{\bf r}\varphi_0({\bf r})\hat{\mit\Psi}
^\dag({\bf r})
\label{c0}
\end{equation}
is the creation operator for atoms in the condensate state
$\varphi_0({\bf r})$, which in the absence of atom-atom interactions
is the ground state of the potential
$V({\bf r})$.

The Heisenberg equation of motion for the condensate field operator
is readily derived from (\ref{H}), giving
\begin{eqnarray}
\frac{d}{dt}\hat{c}_0&=&-i\omega_0\hat{c}_0
-i\frac{\chi}{\sqrt{N}}\int d^3{\bf r}\varphi^\ast_0({\bf r})
\left[\hat{a}^\dag e^{-i{\bf K}\cdot{\bf r}}\right. \nonumber\\
&+&\left. \hat{a}e^{i{\bf K}\cdot{\bf r}}\right]\hat{\mit\Psi}({\bf r}),
\label{dc0dt}
\end{eqnarray}
where $\hbar\omega_0$ is the ground state energy,
$\chi=|g_1||g_2||a_2|\sqrt{N}/|\Delta|$
is the effective coupling constant between the condensate and the probe
field,
and $\hat{a}=(g_1g^\ast_2a^\ast_2\Delta/|g_1||g_2||a_2||\Delta|)\hat{a}_1$
is simply the probe annihilation operator times a phase factor related to
the phase of the pump laser and the sign of the detuning.

\subsection{Free-propagation regime}
From Eq.~(\ref{dc0dt}) we see that the condensate mode is optically coupled
to two new states whose field operators are given by
\begin{equation}
\hat{c}_\pm=\int d^3 {\bf r}\varphi^\ast_0({\bf r})e^{\mp i{\bf K}\cdot{\bf r}}
\hat{\mit\Psi}({\bf r}).
\label{defcpm}
\end{equation}
These states retain the spatial probability distribution of the
condensate state, but propagate at the recoil velocity $v_r=\hbar|{\bf
K}|/m$,
hence we refer to them as `momentum side modes' of the original condensate.
With these definitions, Eq.~(\ref{dc0dt}) becomes
\begin{equation}
\frac{d}{dt}\hat{c}_0=-i\omega_0\hat{c}_0
-i\frac{\chi}{\sqrt{N}}\left[\hat{a}^\dag\hat{c}_++\hat{a}\hat{c}_-\right].
\label{dc0dtfree}
\end{equation}
Due to the fact that the recoil velocity is extremely slow ($\sim 1$ cm/s),
the propagation of the side mode wavepackets
can be neglected for reasonably long times, in which case the
side mode field operators obey
\begin{equation}
\frac{d}{dt}\hat{c}_-=-i(\omega_0+\omega_r)\hat{c}_-
-i\frac{\chi}{\sqrt{N}}\hat{a}^\dag\hat{c}_0,
\label{dcmdtfree}
\end{equation}
and
\begin{equation}
\frac{d}{dt}\hat{c}_+=-i(\omega_0+\omega_r)\hat{c}_+
-i\frac{\chi}{\sqrt{N}}\hat{a}\hat{c}_0,
\label{dcpdtfree}
\end{equation}
where $\omega_r=\hbar|{\bf K}|^2/2m$ is the recoil frequency.
We note that we have neglected coupling to higher order side modes, an
approximation valid in the small-signal regime. Lastly, we note that
in this free propagation model the probe field operator satisfies
the equation of motion
\begin{equation}
\frac{d}{dt}\hat{a}=-i\delta\hat{a}
-i\frac{\chi}{\sqrt{N}}\left[\hat{c}^\dag_-\hat{c}_0+\hat{c}^\dag_0\hat{c}_+
\right],
\label{dadtfree}
\end{equation}
where $\delta=\omega_1-\omega_2$ is the pump-probe detuning,
and we again neglect higher-order momentum side modes.

\subsection{Cavity regime}
The free propagation model is valid for times short enough that atoms at the
recoil velocity propagate only over distances which are short compared
to the dimensions of the initial condensate. As we have shown, in this
regime
the atomic field can be expanded onto momentum side modes, which are simply
plane waves with a slowly varying spatial envelope. A second well-defined
regime
occurs at much longer time scales, when the recoil velocity atoms propagate
over distances which are larger than the trap dimensions. In this
`cavity atom optics' regime, the atomic field operator is best expressed in
terms of the trap eigenmodes $\{\varphi_n({\bf r})\}$ according to
\begin{equation}
\hat{\mit\Psi}({\bf r})=\sum_{n=0}^\infty \varphi_n({\bf r})\hat{c}_n,
\label{modeexpand}
\end{equation}
where $\hat{c}_n$ is the annihilation operator for atoms in mode $n$. With 
this expansion, the equation of motion for the condensate field operator
becomes
\begin{equation}
\frac{d}{dt}\hat{c}_0=-i\omega_0\hat{c}_0
-i\frac{\chi}{\sqrt{N}}\sum_{n=0}^\infty
A_{0n}\left[\hat{a}^\dag+\hat{a}\right]
\hat{c}_n,
\label{dc0dtcav}
\end{equation}
where
\begin{equation}
A_{nn^\prime}=\int d^3{\bf r}\varphi^\ast_n({\bf r})e^{i{\bf K}\cdot{\bf r}}
\varphi_{n^\prime}({\bf r})
\label{A}
\end{equation}
is the matrix element for the optical transition, and we have assumed without
loss of generality that $A_{nn^\prime}$ is a real number.

We assume for simplicity that the matrix elements $A_{nn^\prime}$ is sharply
peaked, so that $|n-n^\prime|=m$ where $m$ is some number. This could be
the case, e.g. in a Fabry-P\'erot-type matter-wave resonator, where the
absolute value of the momentum is relatively well defined for each trap
energy level. This allows us to neglect all excited states except $n=m$,
which gives
\begin{equation}
\frac{d}{dt}\hat{c}_0=-i\omega_0\hat{c}_0
-i\frac{\chi}{\sqrt{N}} A_{0m}\left[\hat{a}^\dag+\hat{a}\right] \hat{c}_m.
\label{dc0dtcav2}
\end{equation}
In this simplified model, the condensate mode is coupled to a single excited
trap level, whose equation of motion is given by
\begin{equation}
\frac{d}{dt}\hat{c}_m=-i\omega_m\hat{c}_m
-i\frac{\chi}{\sqrt{N}} A_{0m}\left[\hat{a}^\dag+\hat{a}\right]\hat{c}_0,
\label{dcmdtcav}
\end{equation}
where $\hbar\omega_m$ is the energy of the
the $m$th trap level, and we again neglect higher-order couplings.
To complete this model, we also require the equation for the probe field
operator,
\begin{equation}
\frac{d}{dt}\hat{a}=-i\delta\hat{a}-i\frac{\chi}{\sqrt{N}} A_{0m}
\left[\hat{c}^\dag_m\hat{c}_0+\hat{c}^\dag_0\hat{c}_m\right].
\label{dadtcav}
\end{equation}
Equations (\ref{dc0dtcav2}-\ref{dadtcav}) differ from
(\ref{dc0dtfree}-\ref{dadtfree}) in that the two running wavepacket
momentum side modes have now been combined into a single standing wave
atomic field mode. In the free-propagation regime, which state the atom
jumps into depends on whether a probe photon was absorbed or emitted,
a result of momentum conservation. In contrast, in the cavity regime both
processes transfer atoms to the same quantum state, due to the uncertainty
in the momentum direction of the trap levels. We will see shortly that
this distinction can lead to remarkably different physical effects.
A similar situation is familiar in conventional atom optics, where the
diffraction of an atom by a standing wave is in general different from
the diffraction by two counterpropagating running waves. \cite{ShoMeySte91}

\subsection{Linear response}

The condensate mode is assumed to be initially occupied by a very large number
$N$ of atoms, in which case it can be treated classically. Introducing a
spontaneous
symmetry breaking ansatz and replacing the operator $\hat{c}_0$ with the
c-number order parameter $c(t)$, and considering furthermore
times short enough that the side mode populations are small compared
to $N$, so that we can neglect condensate depletion, we have
\begin{equation}
c_0(t)=\sqrt{N}e^{-i\omega_0t}.
\label{c0oft}
\end{equation}
It is natural to choose the zero of energy to coincide with
the ground state energy, so that $\omega_0=0$ without loss of
generality. Substituting expression (\ref{c0oft}) into
Eqs.~(\ref{dcmdtfree}-\ref{dadtfree}) and
linearizing in the side mode and probe field operators yields a
closed $3\times 3$ system of equations,
\begin{equation}
\frac{d}{dt}\left(\begin{array}{c}
\hat{c}^\dag_-\\
\hat{c}_+\\
\hat{a}\\
\end{array}\right)=i\left(\begin{array}{ccc}
\omega_r&0&\chi\\
0&-\omega_r&-\chi\\
-\chi&-\chi&-\delta\\
\end{array}\right)\left(\begin{array}{c}
\hat{c}^\dag_-\\
\hat{c}_+\\
\hat{a}\\
\end{array}\right)
\label{3x3free}
\end{equation}
which describes the linear response of the system in the free propagation
regime.
We note that while there are six independent operators $\hat{c}^\dag_-$,
$\hat{c}_+$, $\hat{a}$, and their Hermitian conjugates, the linearized
equations
have decoupled into two sets of three coupled equations due to the momentum
selection rules.

In the cavity regime, there are only four coupled operators,
$\hat{c}_m$, $\hat{c}^\dag_m$, $\hat{a}$ and $\hat{a}^\dag$. The equations
of motion, however, do not decouple and the dynamics is determined by the
$4\times 4$ system of equations
\begin{equation}
\frac{d}{dt}\left(\begin{array}{c}
\hat{c}_m\\ \hat{c}^\dag_m\\ \hat{a}\\ \hat{a}^\dag\\
\end{array}\right)=i\left(\begin{array}{cccc}
-\omega_m&0&-\chi_m&-\chi_m\\
0&\omega_m&\chi_m&\chi_m\\
-\chi_m&-\chi_m&-\delta&0\\
\chi_m&\chi_m&0&\delta\\
\end{array}\right)
\left(\begin{array}{c}
\hat{c}_m\\ \hat{c}^\dag_m\\ \hat{a}\\ \hat{a}^\dag\\
\end{array}\right),
\label{4x4cav}
\end{equation}
where $\chi_m=\chi A_{0m}$.

Equation (\ref{3x3free}) for the free propagation regime, and
Eq.~(\ref{4x4cav}) for the cavity regime can be derived from
the effective Hamiltonians
\begin{eqnarray}
\hat{H}_{free}&=&\hbar\omega_r\left(\hat{c}^\dag_-\hat{c}_-
+\hat{c}^\dag_+\hat{c}_+\right)+\hbar\delta\hat{a}^\dag\hat{a}\nonumber\\
&+&\hbar\chi\left(\hat{a}^\dag\hat{c}^\dag_-+\hat{a}^\dag\hat{c}_+
+\hat{c}^\dag_+\hat{a}+\hat{c}_-\hat{a}\right),
\label{Hfree}
\end{eqnarray}
and
\begin{eqnarray}
\hat{H}_{cav}&=&\hbar\omega_m\hat{c}^\dag_m\hat{c}_m
+\hbar\delta\hat{a}^\dag\hat{a}\nonumber\\
&+&\hbar\chi_m\left(\hat{a}^\dag\hat{c}^\dag_m +\hat{a}^\dag\hat{c}_m
+\hat{c}^\dag_m\hat{a}+\hat{c}_m\hat{a}\right),
\label{Hcav}
\end{eqnarray}
respectively. Terms like $\hat{a}^\dag\hat{c}^\dag$ correspond to the
generation of correlated atom-photon pairs. This is analogous to the optical
parametric amplifier (OPA), except that in that case it is correlated
photon pairs that are generated. The OPA is currently the primary device for
the generation of entangled and/or nonclassical states, and plays a role in
many fundamental experiments in quantum physics, such as tests of Bell's
inequality, quantum cryptography, and quantum teleportation.
This analogy, and the possibility of generating entanglement between atomic
and optical fields is discussed in detail in \cite{MooMey99b,MooMey99}.

Both linear systems can be solved
analytically, all that is required are the eigenvalues and eigenvectors
of the $3\times 3$  and $4\times 4$ matrices which
appear on the right-hand-side of equations (\ref{3x3free}) and
(\ref{4x4cav})
respectively.
The characteristic frequencies in the free space regime satisfy the
cubic equation
\begin{equation}
\omega^3+\delta\omega^2-\omega_r^2\omega-\omega^2_r\delta+2\omega_r\chi^2=0.
\label{cubic}
\end{equation}
This equation either has three real solutions,
in which case the system is stable and experiences only small oscillations
about the initial state, or it has one real and a pair of complex
conjugate solutions. In the latter case the eigenvalue with the negative
imaginary part corresponds to an exponentially growing solution, and hence
the
system is unstable. In this case any small input, even quantum fluctuations,
results in a large output, i.e. the system acts as an amplifier for the
optical
probe and atomic side mode fields. 

In the cavity regime, the characteristic equation is a quartic equation,
given by
\begin{equation}
\omega^4-(\delta^2+\omega_m^2)\omega^2+\delta^2\omega_m^2
-4\delta\omega_m\chi_m^2=0,
\label{quartic}
\end{equation}
which is simply a quadratic equation in $\omega^2$. The solutions to
Eq.~(\ref{quartic}) fall into three categories. The first category is of
course
when all solutions are purely real, in which case the system is stable.
A second possibility is that there
are two purely real and two purely imaginary solutions of the form
$\{\omega_1=\Omega, \omega_2=-\Omega, \omega_3=i\Gamma, \omega_4=-i\Gamma\},$
where $\Omega$ and $\Gamma$ are both real quantities. In this case there is
only
one exponentially growing solution, at the imaginary frequency $\omega_4$.
A final possibility is that the solutions are complex numbers of the form
$\{\omega_1=\Omega+i\Gamma, \omega_2=-\Omega+i\Gamma,
\omega_3=\Omega-i\Gamma,
\omega_4=-\Omega-i\Gamma\}$. This last case is interesting in that there
are two exponentially growing solutions, $\omega_3$ and $\omega_4$,
which grow at the same rate $\Gamma$, but rotate at equal and opposite
frequencies $\pm\Omega$. This leads to the temporal modulation of the 
exponential growth rate.

In Fig.~1 we plot the domains of parameter space corresponding to the 
various types of solutions. The shaded areas correspond to
the unstable regions where exponential growth occurs.
Figure 1a shows the free propagation model, where
the stable and unstable regions are depicted in the $\delta$-$\chi^2$
plane. Both $\delta$ and $\chi$ are taken in units of $\omega_r$.
In Fig 1b, we plot the corresponding results for the cavity model
in the $\delta$-$\chi^2_m$ plane. Here $\delta$ and $\chi_m$ are taken in units of
$\omega_m$. In the figures, roman numeral I corresponds
to a single exponentially growing solution and roman numeral II corresponds
to two exponentially growing solutions. 

\section{mean intensities}

In this section we use the linearized description of section II to derive 
expressions for the mean intensities of the probe and side mode fields, and 
compare the results for the two models. We focus in particular on the 
`exponential growth regime', which occurs for times long enough that all
but the exponentially growing solutions of section II can be safely
neglected,
yet short enough that condensate depletion can be ignored. We assume for
concreteness that the atomic side modes begin in the vacuum state, while
the probe field is initially in a coherent state $\alpha$, corresponding to
the
injection of a weak laser field into the ring cavity.

\subsection{Free propagation regime}

In the free propagation regime, the solutions to Eq.~(\ref{3x3free}) is
given by
\begin{equation}
\left(\begin{array}{c}
\hat{c}^\dag_-(t)\\ \hat{c}_+(t) \\ \hat{a}(t)\\
\end{array}\right)=
{\bf A}(t)
\left(\begin{array}{c}
\hat{c}^\dag_-(0)\\ \hat{c}_+(0) \\ \hat{a}(0)\\
\end{array}\right).
\label{solutionfree}
\end{equation}
The $3\times 3$ matrix ${\bf A}(t)$ is given by
\begin{equation}
{\bf A}(t)={\bf U}\exp(i{\bf W} t){\bf U}^{-1},
\label{defAoft}
\end{equation}
where $U_{ik}$ is the $i$th component of the $k$th eigenvector of
the matrix on the right-hand-side of Eq.~(\ref{3x3free}),
and ${\bf W}$ is the corresponding diagonal matrix of eigenvalues.
For times long compared to the exponential doubling rate, we
can neglect all but the exponentially growing terms,
in which case we have
\begin{equation}
A_{ij}(t)\approx
a_{ij}e^{i\Omega+\Gamma t},
\label{defa}
\end{equation}
where $a_{ij}$ is a constant which depend only on the
control parameters $\delta$ and $\chi$.

From expression (\ref{solutionfree}),
together with the specified initial condition, we find that
the mean intensities $I_1=\langle\hat{c}^\dag_-\hat{c}_-\rangle$,
$I_2=\langle\hat{c}^\dag_+\hat{c}_+\rangle$, and
$I_3=\langle\hat{a}^\dag\hat{a}\rangle$ take the form $I_j(t)=\tilde{I}_j
\exp(2\Gamma t)$, where
\begin{equation}
\tilde{I}_j=|a_{j1}|^2+|\alpha|^2|a_{j3}|^2.
\label{intensitiesfree}
\end{equation}
This expression contains a spontaneous contribution,
present even in the case $\alpha=0$, as well as a stimulated
term proportional to the injected signal strength $|\alpha|^2$.
The former is due to the amplification of vacuum fluctuations in the
atomic density and the probe electromagnetic field, whereas the latter
is due to amplification of the injected probe field.

\subsection{Cavity atom optics regime}

In the cavity regime, the solution to Eq.~(\ref{4x4cav})
is given by
\begin{equation}
\left(\begin{array}{c}
\hat{c}_m(t)\\ \hat{c}^\dag_m(t)\\ \hat{a}(t)\\ \hat{a}^\dag(t)\\
\end{array}\right)=
{\bf B}(t)
\left(\begin{array}{c}
\hat{c}_m(0)\\ \hat{c}^\dag_m(0)\\ \hat{a}(0)\\ \hat{a}^\dag(0)\\
\end{array}\right),
\label{solutioncavI}
\end{equation}
where the $4\times 4$ matrix ${\bf B}$ is defined in the same manner as
${\bf
A}$, but the eigenvalues and eigenvectors are those of the $4\times 4$
matrix on
the right-hand-side of Eq.~(\ref{4x4cav}).

Again for moderately long times
we keep only exponentially growing terms, which in region I of Fig.~1b
allows us to take
\begin{equation}
B_{ij}(t)\approx b_{ij}e^{\Gamma t},
\label{defb}
\end{equation}
where
$b_{ij}$ is a constant which depends only on the parameters
$\delta$ and $\chi_m$.
 With this approximation, we find that the mean intensities,
$I_1=\langle\hat{c}^\dag_m\hat{c}_m\rangle$ and
$I_3=\langle\hat{a}^\dag\hat{a}\rangle$, are all of the form
$I_j(t)=\tilde{I}_j\exp(2\Gamma t)$, where
\begin{eqnarray}
\tilde{I}_j&=&|b_{j2}|^2+|b_{j4}|^2
+|\alpha|^2\left(|b_{j3}|^2+|b_{j4}|^2\right)\nonumber\\
&+&2{\mbox Re}\left[b_{j3}b^\ast_{j4}\alpha^2\right] .
\label{intensitiescavI}
\end{eqnarray}
As in the free propagation model, there is a spontaneous component, present
even for $\alpha=0$, and a stimulated component, which is present only for
$\alpha \ne 0$. One important difference, however, is that in contrast to
the free propagation regime of Eq.~(\ref{intensitiesfree}),
where the intensities depend only on $|\alpha|$, in the cavity situation 
the intensities are also sensitive to the
phase of the initial probe field. We remark that the intensities in
region I of the cavity model and those
in the free propagation model are very similar in that the time-dependence
is a pure exponential. This is no longer the case in
unstable region II of the cavity model.

In region II of Fig.~1b, there are two independent exponentially growing
solutions. Keeping only these terms gives
\begin{equation}
B_{ij}(t)\approx\left(c_{ij}e^{i\Omega t}+d_{ij}e^{-i\Omega t}
\right)e^{\Gamma t},
\label{defbpm}
\end{equation}
where $c_{ij}$ and $d_{ij}$ are again constants which depend only on the
parameters $\delta$ and $\chi_m$.
Due to the counterrotating terms, one immediately
expects to see oscillations, or beating, superimposed onto the exponential
growth.

The mean intensities in region II take the form
$I_j(t)=\tilde{I}_j(t)\exp(2\Gamma t)$, where
\begin{eqnarray}
\tilde{I}_j(t)&=&
X_j+2{\mbox Re}\left[x_je^{i2\Omega t}\right]
+|\alpha|^2\left(Y_j
+2{\mbox Re}\left[y_je^{i2\Omega t}\right]\right)\nonumber\\
&+&2{\mbox Re}\left[\left(Z_j+z_je^{i2\Omega t}
+\bar{z}_je^{-i2\Omega t}\right)\alpha^2\right],
\label{intensitiescavII}
\end{eqnarray}
where the various parameters are
\begin{equation}
X_j=|c_{j2}|^2+|d_{j2}|^2+|c_{j4}|^2+|d_{j4}|^2,
\label{Xj}
\end{equation}
\begin{equation}
x_j=c_{j2}d^\ast_{j2}+c_{j4}d^\ast_{j4},
\label{xj}
\end{equation}
\begin{equation}
Y_j=|c_{j3}|^2+|d_{j3}|^2+|c_{j4}|^2+|d_{j4}|^2,
\label{Yj}
\end{equation}
\begin{equation}
y_j=c_{j3}d^\ast_{j3}+c_{j4}d^\ast_{j4},
\label{yj}
\end{equation}
\begin{equation}
Z_j=c_{j3}c^\ast_{j4}+d_{j3}d^\ast_{j4},
\label{Zj}
\end{equation}
\begin{equation}
z_j=c_{j3}d^\ast_{j4},
\label{zj}
\end{equation}
and
\begin{equation}
\bar{z}_j=d_{j3}c^\ast_{j4}.
\label{barzj}
\end{equation}
From Eq.~(\ref{intensitiescavII}) we see that in addition to purely
exponential terms, there are now contributions with superimposed
oscillations. These beats occur even when quantum noise alone is amplified,
as can be seen in the first and second terms in Eq.~(\ref{intensitiescavII}).
These terms are spontaneous contributions, present even in the absence of an 
injected probe field $(\alpha=0)$. In contrast, the third term is a
stimulated
contribution which is independent of the phase of the injected field and
the fourth term, which gives additional oscillatory and non-oscillatory
terms,
is sensitive to the phase of the injected field $\alpha$.

We remark that when the transcendental equation
\begin{equation}
Z_j+z_je^{i2\Omega t}
+\bar{z}_je^{-i2\Omega t}=0
\label{transcendental}
\end{equation}
appearing in that last contribution is satisfied, say at a time $t=t_0$,
then the phase sensitivity of the solution disappears. This property recurs
for the times $t=t_0+n\pi/\Omega$, where $n$ is any integer. At these times,
the intensities corresponding to initial states having
the same $|\alpha|$ but different phases intersect.

The growth of the matter-wave intensity is illustrated in Fig.~2 for two
examples. Figure 2a shows $\ln|I_3|$ versus $t$
for the case $\alpha=1$ for $\chi_m=\omega_m^{-1}$ and
$\delta=-\omega_m^{-1}$, 
a set of system parameters in region I. Here, there is only one
exponentially 
growing solution and no oscillations. Figure 2b shows the result for
$\chi_m=\omega_m^{-1}$ and $\delta=\omega_m^{-1}$. This example demonstrates
the beating which occurs in region II as a result of the interplay between
the two counterrotating exponential terms. In both figures $t$ is taken to be
in units of $\omega_m$.

\section{Summary and Conclusion}

The fact that Bose-Einstein condensates of low density atomic vapors
are generated in traps has important consequences on their intrinsic
properties.
For instance, in a harmonic trap the dependence of the number of condensate
atoms on the critical temperature scales as $1-(T/T_c)^3$, rather than the
free-space result $1 -(T/T_c)^{3/2}$. The condensate ground state and
the spectrum of quasi-excitation also depends drastically on the presence
and form of a trap potential, so much so that it is even possible to create
stable small trapped condensates of atoms subject to an attractive
interaction.

In this paper, we have discussed how the trap environment also strongly
influences the way it interacts with light. We have shown that after times
long enough that the condensate atoms can probe their environment, the way
condensate side modes and light fields are amplified differs significantly
from the free-space situation. The situation is analogous to cavity
QED in quantum optics, except that in the present case, one tailors the
matter-wave mode density rather than the density of modes of the optical
field.  

We have considered the example of the ultracold collective
atom laser, or ``free-atom laser,'' which is an atomic matter-wave analog of
the free-electron laser. We have demonstrated how the stimulated scattering 
of light by matter-wave fields leading to their mutual amplification is 
drastically changed in matter-wave resonators, as compared to the free-space
situation. This paper has concentrated on the growth rate of the intensity
of the side modes of both the condensate and the optical field. Further
work will investigate in which way such trap environments also modify the
quantum statistics of these fields and their entanglement.

\acknowledgements
This work is supported in part by the U.S. Office of Naval Research
under Contract No.\ 14-91-J1205, by the National Science Foundation under
Grant No.\ PHY98-01099, by the U.S.\ Army Research Office, and by the
Joint Services Optics Program.

This paper is dedicated to Marlan Scully on his 60th Birthday.\

\begin{figure}
\caption{Exponential instability regions for the free propagation and
cavity regimes are shown in Figs. 1a and 1b respectively.  Region I corresponds
to one exponentially growing solution, whereas in region II there are two
counterrotating exponential solutions. In Fig.~1a, $\delta$ and $\chi$ are
taken in units of $\omega_r$, while in Fig.~1b $\delta$ and $\chi_m$
are in units of $\omega_m$.}
\end{figure}
\begin{figure}
\caption{Figure 2a shows $\ln|I_3|$ versus $t$
for the case $\alpha=1$ for $\chi_m=\omega_m^{-1}$ and
$\delta=-\omega_m^{-1}$, 
a set of system parameters in region I. Figure 2b shows the result for
$\chi_m=\omega_m^{-1}$ and $\delta=\omega_m^{-1}$. This example demonstrates
the beating which occurs in region II as a result of the interplay between
the two counterrotating exponential terms. In both figures $t$ is taken to be
in units of $\omega_m$.}
\end{figure} 

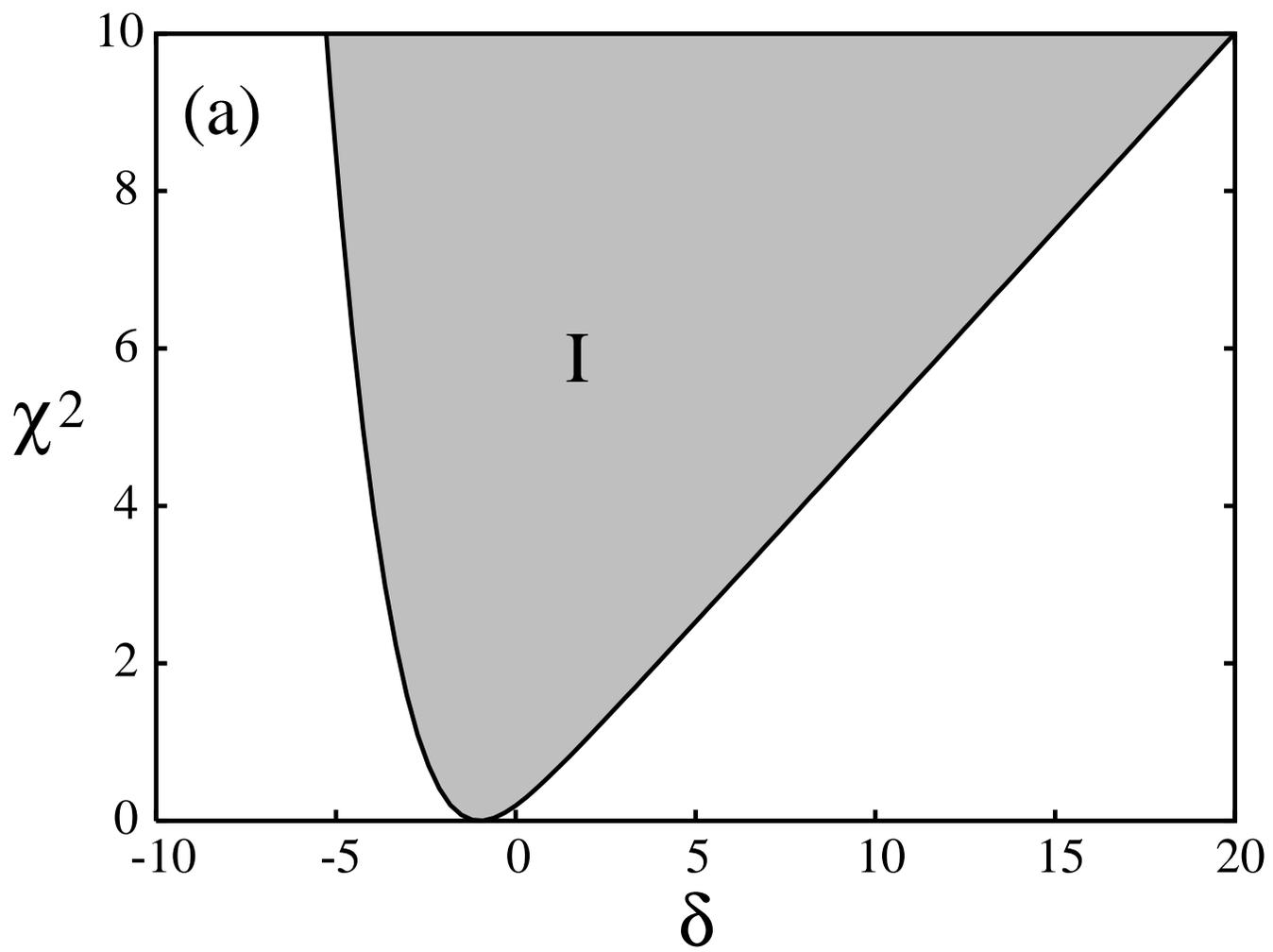
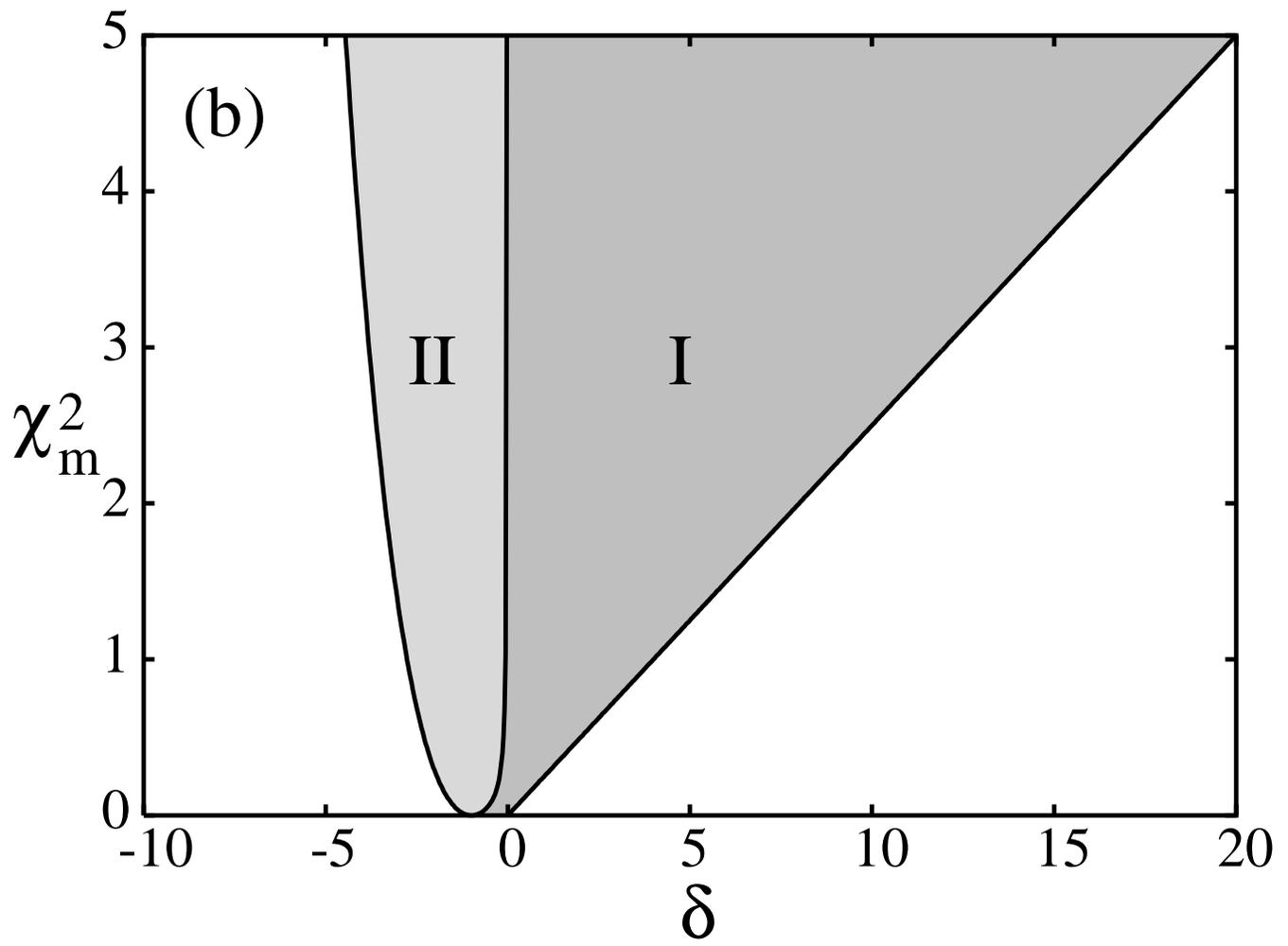
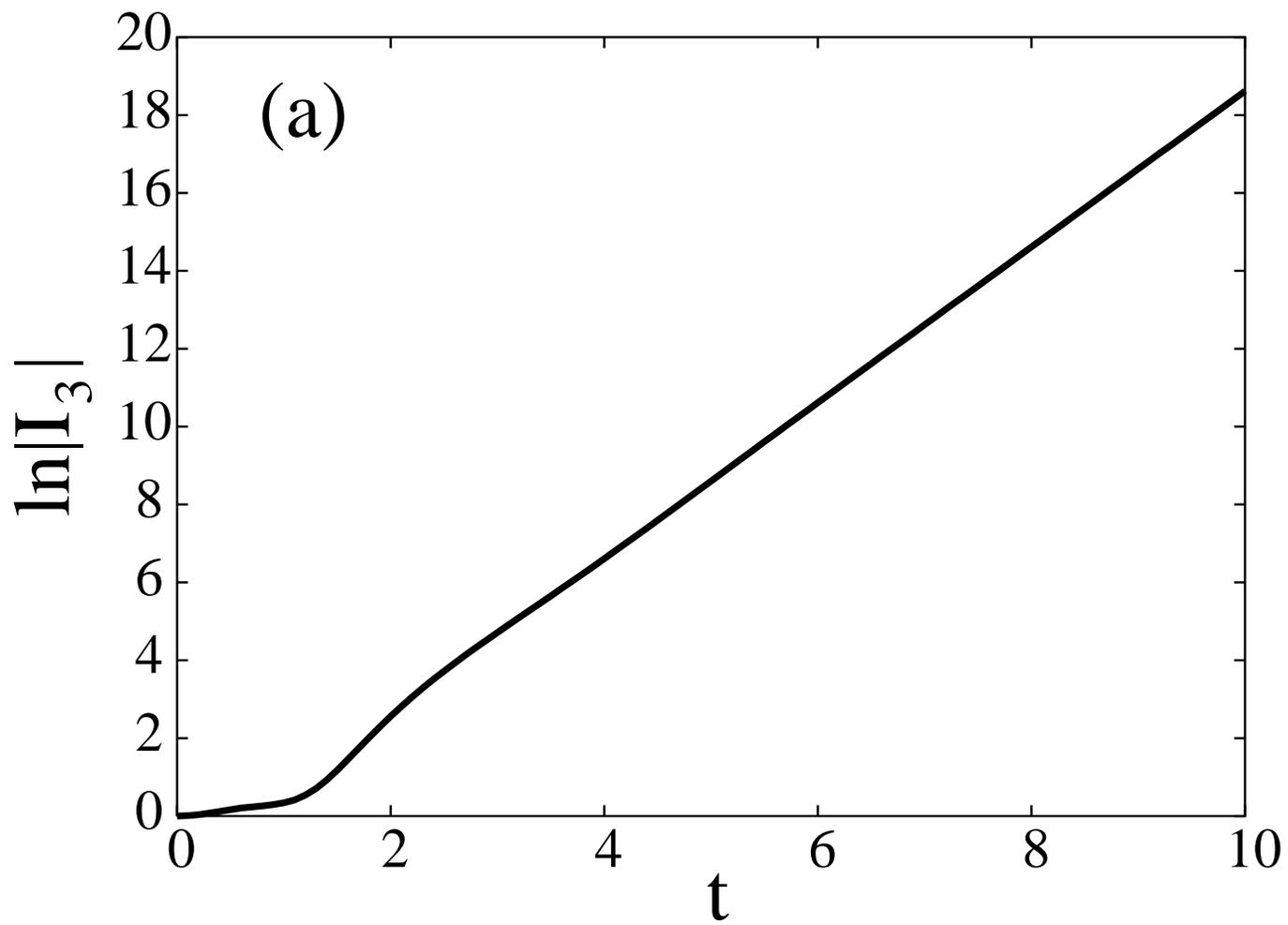
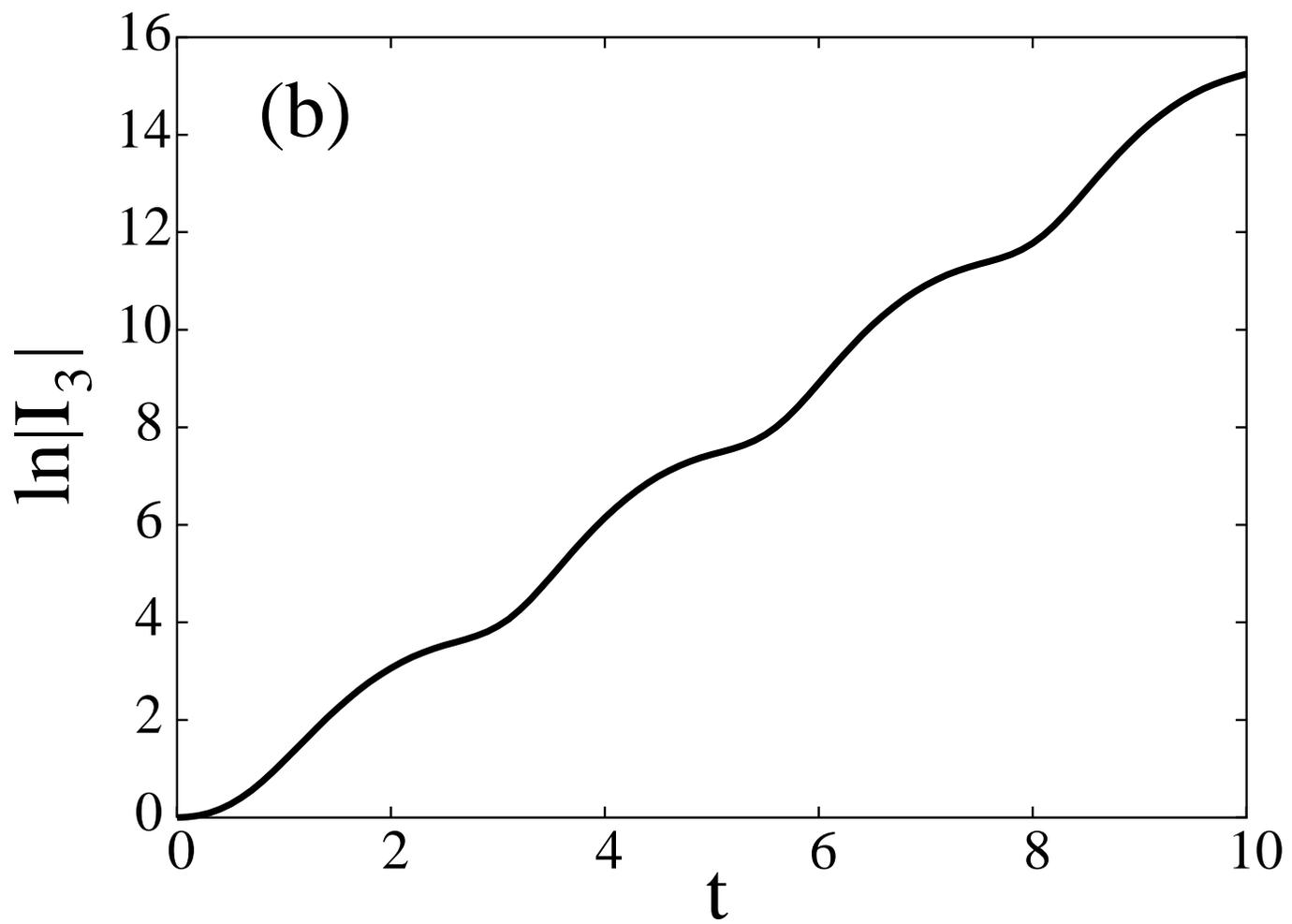
\end{document}